%% file: Final.tex
\documentclass[journal,twocolumn,10pt]{IEEEtran}
\input{preamble}
\IEEEoverridecommandlockouts
\begin{document}

\title{Digital and Hybrid Precoding Designs in Massive MIMO with Low-Resolution ADCs}

\author{Mengyuan~Ma,~\IEEEmembership{Student Member,~IEEE}, Nhan~Thanh~Nguyen,~\IEEEmembership{Member,~IEEE},\\ Italo~Atzeni,~\IEEEmembership{Senior~Member,~IEEE}, 
~A.~Lee~Swindlehurst,~\IEEEmembership{Fellow,~IEEE},
and~Markku~Juntti,~\IEEEmembership{Fellow,~IEEE}

\thanks{This work was supported by the Research Council of Finland (332362 EERA, 354901 DIRECTION, 336449 Profi6, 348396 HIGH-6G, 357504 EETCAMD, and 369116 6G~Flagship), by EU CHSIT-ERA (359817 PASSIONATE), and by the U.S. National Science Foundation (CCF-2225575). M.~Ma, N.~T.~Nguyen, I.~Atzeni, and M.~Juntti are with the Centre for Wireless Communications, University of Oulu, Finland (e-mail: \{mengyuan.ma, nhan.nguyen, markku.juntti\}@oulu.fi, italo.atzeni@ieee.org). A.~L.~Swindlehurst is with the with the Center for Pervasive Communications \& Computing, University of California, Irvine, CA, USA (e-mail: swindle@uci.edu).}

\vspace{-5mm}

}

\maketitle

\begin{abstract}
Low-resolution analog-to-digital converters (ADCs) have emerged as an efficient solution for massive multiple-input multiple-output (MIMO) systems to reap high data rates with reasonable power consumption and hardware complexity. In this paper, we study precoding designs for digital, fully connected (FC) hybrid, and partially connected (PC) hybrid beamforming architectures in massive MIMO systems with low-resolution ADCs at the receiver. We aim to maximize the spectral efficiency (SE) subject to a transmit power budget and hardware constraints on the analog components. The resulting problems are nonconvex and the quantization distortion introduces additional challenges. To address them, we first derive a tight lower bound for the SE, based on which we optimize the precoders for the three beamforming architectures under the majorization-minorization framework. Numerical results validate the superiority of the proposed precoding designs over their state-of-the-art counterparts in systems with low-resolution ADCs, particularly those with 1-bit resolution. The results show that the PC hybrid precoding design can achieve an SE close to those of the digital and FC hybrid precoding designs in 1-bit systems, highlighting the potential of the PC hybrid beamforming architectures.
\end{abstract}

\begin{IEEEkeywords}
Digital precoding, hybrid precoding, low-resolution ADCs, massive MIMO.
\end{IEEEkeywords}

%
\IEEEpeerreviewmaketitle

\medskip

\section{Introduction}\label{sec: introduction}

 Massive multiple-input multiple-output (MIMO) and beamforming technologies are crucial for wireless communications at both sub-6~GHz and millimeter-wave (mmWave) frequencies, enabling a high spectral efficiency (SE) \cite{bjornson2019massive}. However, a large number of power-hungry radio-frequency (RF) chains would cause high power consumption, degrading the system's energy efficiency (EE). In this respect, analog-to-digital converters (ADCs) are the most power-consuming RF components at the receiver, as their power consumption increases exponentially with the number of resolution bits \cite{murmann2015race}. Therefore, efficient beamforming designs with low-resolution ADCs are promising to reduce the power consumption without excessively compromising the SE performance \cite{Atz22}. 

Existing precoding algorithms (e.g., \cite{yu2016alternating,gao2016energy,nguyen2019unequally,ma2021closed}) are mainly designed for digital beamforming (DBF) and hybrid beamforming (HBF) architectures. DBF architectures require a dedicated RF chain for each antenna, providing large spatial multiplexing gains at the expense of high power consumption. In contrast, HBF architectures employ fewer RF chains with a network of analog components, such as phase shifters, to cut the hardware power consumption and cost at the expense of reduced spatial multiplexing. HBF implementations can be further divided into fully connected (FC) and partially connected (PC) HBF (FC-HBF and PC-HBF) architectures. In the former, each RF chain can access all the antenna elements, while in the latter only a subset of the antenna elements is connected to each RF chain. 

Traditional FC and PC hybrid precoding designs \cite{yu2016alternating,gao2016energy,nguyen2019unequally,ma2021closed} have predominantly focused on full-resolution systems, i.e., with high-resolution ADCs at the receiver. However, these designs tend to be less effective in systems with low-resolution ADCs since they do not take the quantization distortion (QD) at the receiver into account. To address this challenge, the works in \cite{mo2017hybrid,lin2016energy,hou2018hybrid} employed a heuristic two-step approach that first obtains a feasible analog precoder and then optimizes the digital precoder based on the analog one. For instance, Mo {\it et. al.} \cite{mo2017hybrid} derived the analog precoder based on the singular vectors of the channel, followed by the water-filling (WF) method to optimize the digital part. In \cite{lin2016energy}, the analog precoder was chosen from a predefined codebook, after which the digital precoder was optimized to maximize the EE. Beyond HBF architectures, the QD introduces further challenges for the digital precoding design. Conventional precoding strategies, including zero forcing, maximum ratio transmission, and minimum mean squared error, were explored in \cite{zhang2021analysis,usman2016mmse,lin2022learning}. In  \cite{ma2024joint}, we jointly optimized the digital precoder and combiner, resulting in a substantial SE improvement over traditional WF solutions.

Despite the advances discussed above \cite{mo2017hybrid,lin2016energy,hou2018hybrid,zhang2021analysis,usman2016mmse,lin2022learning,ma2024joint}, most studies have focused on only one of the three beamforming architectures: DBF, FC-HBF, or PC-HBF. Furthermore, HBF studies have often relied on heuristic methods for the analog precoding design, leaving considerable room for improving the SE. To fill this gap, we herein introduce a novel methodology aiming to maximize the SE. Specifically, to tackle the challenges posed by the hardware constraints at the transmitter and the QD at the receiver, we derive a tight lower bound on the SE, based on which we develop iterative precoding algorithms for the DBF, FC-HBF, and PC-HBF architectures using the majorization-minimization (MM) framework. In contrast to prior works, our design guarantees maximization of the SE. Numerical results demonstrate that the proposed precoding designs outperform state-of-the-art methods \cite{gao2016energy,yu2016alternating,mo2017hybrid,ma2024joint} in systems with low-resolution ADCs, particularly those with 1-bit 
resolution. Notably, our results indicate that the PC hybrid precoding design can achieve an SE close to those of the digital and FC hybrid precoding designs in 1-bit systems, highlighting the potential of PC-HBF architectures.

\medskip

\section{System and Quantization Models}\label{eq:DBP design}
\subsection{System Model}
\begin{figure}[t]
\small
    \centering
    \hspace{-3mm}
    \subfigure[FC-HBF.]
    {\label{fig:FC-HBF_receiver} \includegraphics[scale=0.38]{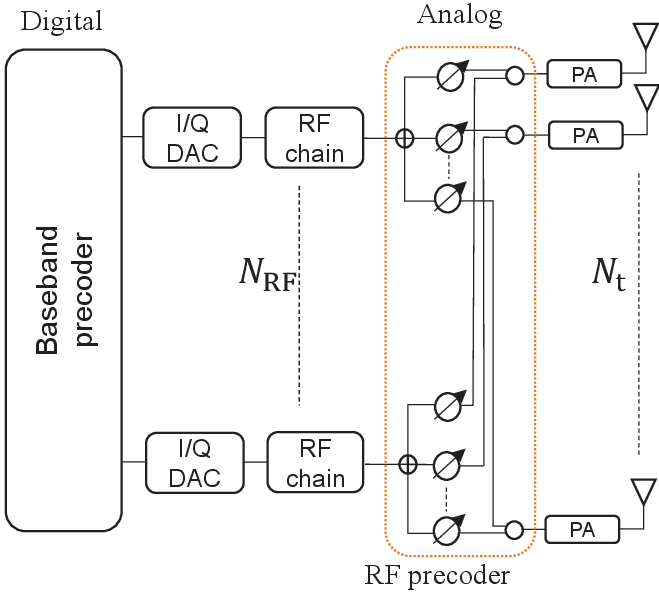}} \hspace{2mm}
        \subfigure[PC-HBF.]
    {\label{fig:PC-HBF_receiver} \includegraphics[scale=0.38]{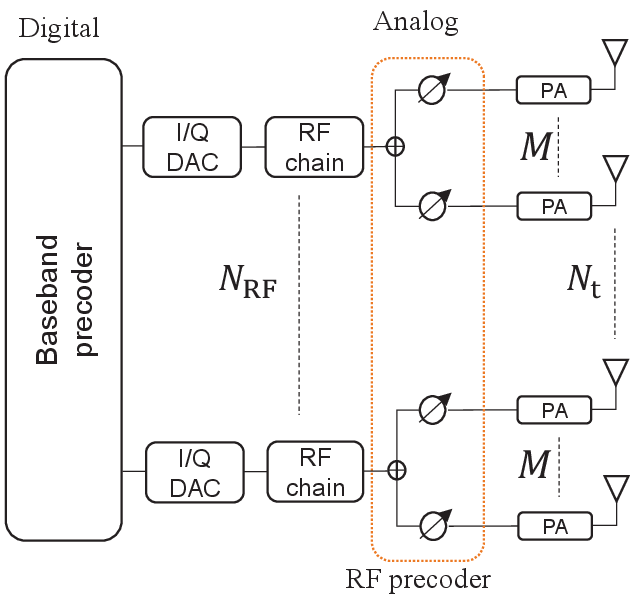}}
    \vspace{-2mm}
    \caption{Illustration of the considered HBF architectures.}
    \label{fig:RX beamforming architectures}
\end{figure}
We consider a point-to-point narrowband MIMO system where a transmitter (Tx) with $\Nt$ antennas sends signals to a digital receiver (Rx) with $\Nr$ antennas, where the latter employs low-resolution ADCs. Let $\ssb \in \Cs^{\Ns} \sim \Ccl\Ncl(0, \Ib)$ denote the transmitted signal vector of $\Ns$ data streams and let $\Fb\in\Cs^{\Nt \times \Ns}$ be the precoding matrix with power constraint $\|\Fb\|_{\rm F}^2 \leq \Pt$, where $\Pt$ denotes the transmit power budget and $\|\cdot\|_{\rm F}$ represents the Frobenius norm. For HBF architectures, $\Fb$ is given by $\Fb=\Frf\Fbb$, where $\Frf \in \Cs^{\Nt\times \Nrf}$ and $\Fbb \in \Cs^{\Nrf\times \Ns}$ are the analog (RF) and digital (baseband) precoders, respectively, assuming $\Nrf$ RF chains ($\Ns\leq \Nrf \ll \Nt $). Fig.~\ref{fig:RX beamforming architectures} illustrates the FC-HBF and PC-HBF architectures. Each RF chain in the FC-HBF architecture is connected to all antennas through a network of phase shifters, while each RF chain in the PC-HBF architecture is connected only to a subset of $M=\frac{\Nt}{\Nrf}$ antennas. For simplicity, we assume that $M$ is an integer and the $\Nrf$ groups of antennas are non-overlapping. Let $\Fcl_{\rm full}$ and $\Fcl_{\rm sub}$ denote the sets of feasible analog precoders for the FC-HBF and PC-HBF architectures, respectively, which can be expressed as
\begin{align}
    \Fcl_{\rm full}&  =\{\Fb_{\rm RF}: |\Fb_{\rm RF}(i,j)|=1,~\forall i,j \}, \label{eq:definition of set for FC-HBF}\\
   \Fcl_{\rm sub} &  = \big\{ \Frf: \Frf=\blkdiag\left(\fb_1,\ldots,\fb_{\rm \Nrf} \right), \nonumber \\
    &\hspace{2cm} \ \fb_n \in \Cs^{M},|\fb_n(i)|=1,~\forall n,i \big\},
\end{align}
where $\Ab(i,j)$ and $\ab(i)$ denote the ($i$,$j$)-th entry of $\Ab$ and the $i$-th element of $\ab$, respectively.  

Let $\Hb \in \Cs^{\Nr\times \Nt}$ be the channel between the Tx and Rx. The received signal can be expressed as
\begin{equation}\label{eq:received signal}
    \yb=\Hb\Fb\ssb + \nb,
\end{equation}
where $\nb \sim \Ccl\Ncl(0, \sigman^2 \Ib)$ denotes an additive white Gaussian noise (AWGN) vector with power $\sigman^2$.  With fixed $\Hb$ and $\Fb$, we have $\yb \sim \Ccl\Ncl(0, \Cb_{\yb})$, where $\Cb_{\yb} =\Es[\yb \yb^\H]=\Hb \Fb \Fb^\H \Hb^\H + \sigman^2 \Ib$ is the covariance matrix of $\yb$.
We assume that the channel is constant during each coherence block and perfect channel state information (CSI) is available at both the Tx and Rx \cite{yu2016alternating,gao2016energy,nguyen2019unequally,mo2017hybrid,lin2016energy,hou2018hybrid,zhang2021analysis,usman2016mmse,lin2022learning,ma2024joint}. The impact of imperfect CSI is evaluated in Section~\ref{sec_simulation}. The quantization model is detailed next.

\subsection{Quantization Model}
Assume that the ADCs are Lloyd-Max quantizers and that the Rx employs two identical ADCs in each RF chain to separately quantize the in-phase and quadrature signals. The codebook for a scalar quantizer of $b$ bits is defined as $\Ccl=\{c_0, \ldots, c_{N_{\rm q}-1}\}$, where $N_{\rm q}=2^b$ is the number of output levels of the quantizer. The set of quantization thresholds is $\Tcl=\{ t_0, \ldots, t_{N_{\rm q}} \}$, where $t_0=-\infty$ and $t_{N_{\rm q}}=\infty$ allow inputs with arbitrary power. Let $Q(\cdot)$ denote the quantization function. For a complex signal $x=\Re\{x\}+j \Im\{x\}$, we have $Q(x)=Q\big(\Re\{x\}\big)+j Q\big(\Im\{x\}\big)$, with $Q\big(\Re\{x\}\big)=c_{i}$ when $\Re\{x\}\in [t_i, t_{i+1}]$; $Q\big(\Im\{x\}\big)$ is obtained similarly.
When the quantizer input is a vector, $Q(\cdot)$ is applied element-wise.

Let $\zb=Q(\yb)$ be the quantization of $\yb$. We model the quantization via the Bussgang decomposition \cite{demir2020bussgang}, yielding the linear representation
\begin{equation}\label{eq:quantized received signal}
	\zb=\Gb\yb + \etab,
\end{equation}
where $\Gb$ and $\etab$ represent the Bussgang gain matrix and the non-Gaussian QD vector, respectively. Note that $\Gb$ ensures that $\etab$ is uncorrelated with $\yb$ \cite{demir2020bussgang},  with $\Gb=\Ib-\Gammab$  and $\Gammab= \diag(\gamma_1,\cdots,\gamma_{\Nr})$. Here $\gamma_i$ denotes the distortion factor of the ADCs for the $i$-th RF chain and can be determined as  $\gamma(b) \approx 2^{-1.74b+0.28}$ for $b$-bit quantization \cite{ma2024joint}. 
Combined with \eqref{eq:received signal}, \eqref{eq:quantized received signal} can be rewritten as
\begin{equation}
    \zb=\Gb\Hb\Fb\ssb+\eb,
\end{equation}
where $\eb = \Gb\nb +\etab$ represents the effective noise, which is not Gaussian due to the presence of the QD. 

\medskip

\section{Digital and Hybrid Precoding Designs}\label{sec:Type-I DBF Design} 
\subsection{Problem Formulation}
Treating the effective noise as Gaussian, we obtain a lower bound for the SE as \cite{mo2017hybrid,ma2024joint}
\begin{equation}\label{eq:achievable lb}
   R=\log \det \left( \Ib +  \Cb_{\eb}^{-1} \Gb \Hb \Fb \Fb^\H\Hb^\H \Gb \right),
\end{equation}
where $\Cb_{\eb}=\Es[\eb \eb^\H]=\Cb_{\etab}+ \sigman^2\Gb^2$ and $\Cb_{\etab} = \Es[\etab \etab^\H]$ are the covariance matrices of $\eb$ and $\etab$, respectively. Note that the covariance matrix of the QD can be approximated as $\Cb_{\etab} \approx \Gb\left(\Ib-\Gb\right)\diag(\Cb_{\yb}) $\cite{ma2024joint},
with which we obtain
\begin{equation}\label{eq:CoV eff approx}
    \Cb_{\eb} \approx \Gb\left(\Ib-\Gb\right)\diag\left(\Hb \Fb \Fb^\H \Hb^\H\right)+ \sigman^2\Gb.
\end{equation}
Note that \eqref{eq:CoV eff approx} takes the QD into consideration. For high-resolution ADCs, $\Gb$ reduces to $\Ib$, which results in the conventional SE expression.
We are interested in optimizing the precoder $\Fb$ to maximize the SE in \eqref{eq:achievable lb}, i.e.,
\begin{equation}\label{pb:DBF design}
   \underset{\|\Fb\|^2_{\rm F}\leq \Pt}{\mathrm{maximize}} \; R(\Fb),
\end{equation}
where $R(\Fb)$ represents the SE achieved by $\Fb$. For the HBF architectures, the additional constraints
\begin{align}
    &\Fb=\Frf \Fbb, \label{eq: coupling constraints}\\
    & \Frf \in \Fcl \label{eq: analog components}
\end{align}
must be satisfied, with $\Fcl=\Fcl_{\rm full}$ for FC-HBF and $\Fcl=\Fcl_{\rm sub}$ for PC-HBF. Problem \eqref{pb:DBF design} is nonconvex and more challenging compared with conventional precoding designs in full-resolution systems because $\Cb_{\eb}$ depends on $\Fb$ through \eqref{eq:CoV eff approx}. To overcome this difficulty, we propose an efficient iterative algorithm based on the MM framework. To make the problem more tractable, we first derive a tight lower bound for $R(\Fb)$ in the following lemma.

\vspace{-4pt}
\begin{lemma}\label{thrm:log-det MM surrogate function}
    Let $\hat{\Fb}$ be a prior estimate of $\Fb$ and define
    \begin{align}
    g(\Fb,\hat{\Fb}) &= R(\hat{\Fb}) -\tr ( \hat{\Xb}^\H \hat{\Cb}_{\eb}^{-1}\hat{\Xb}) + 2\Re\big\{\tr(\hat{\Xb}^\H\hat{\Cb}_{\eb}^{-1}\Xb )\big\} \nonumber\\ 
    &\hspace{3cm} - \tr\big\{ \hat{\Sb} (\Cb_{\eb} +\Xb\Xb^\H)  \big\},
    \end{align}
with $\Xb= \Gb \Hb \Fb$, $\hat{\Cb}_{\eb}=\Gb\left(\Ib-\Gb\right)\diag(\Hb  \hat{\Fb}  \hat{\Fb}^\H \Hb^\H)+ \sigman^2\Gb$, $\hat{\Xb}=\Gb \Hb \hat{\Fb}$, and $\hat{\Sb} =  \hat{\Cb}_{\eb}^{-1}- (\hat{\Cb}_{\eb}+\hat{\Xb}\hat{\Xb}^\H)^{-1}$. It can be shown that $g(\Fb,\hat{\Fb})$ is a tight lower bound for $R(\Fb)$, i.e., $g(\Fb,\hat{\Fb}) \leq R(\Fb)$, where the equality holds only if $\hat{\Fb}=\Fb$.
\end{lemma}

\noindent Lemma~\ref{thrm:log-det MM surrogate function} is obtained based on \cite[Proposition~7]{zhang2021rate}; the proof is omitted due to space limitations. Unlike \cite{zhang2021rate}, which considers full-resolution ADCs, herein we aim to address precoding designs for systems with low-resolution ADCs, where the QD introduces additional challenges for the SE optimization as seen from \eqref{eq:CoV eff approx}. Using Lemma~\ref{thrm:log-det MM surrogate function}, we can update $\Fb$ by iteratively maximizing $g(\Fb,\hat{\Fb})$ using the MM framework. Next, we use this method for digital and hybrid precoding designs.

\subsection{Digital Precoding Design}
 
Let $p_i$ be the power allocated to the $i$-th data stream and define $\pb= [p_1,\ldots,p_{\Ns}]^\T$. 
The precoder can be constructed as 
\begin{equation}\label{eq:SCA precoder}
    \Fb=\Vb\diag (\pb)^{\frac{1}{2}},
\end{equation}
where $\Vb$ contains the $\Ns$ right singular vectors of $\Hb$ associated with the $\Ns$ largest singular values. Thus, the precoding design becomes the power allocation problem $\mathrm{maximize}_{\|\pb\|_1 \leq \Pt}  \, g(\pb, \hat{\pb})$,
following Lemma~\ref{thrm:log-det MM surrogate function}. After some algebraic manipulations, we obtain the equivalent problem
\begin{subequations}\label{pb:simplified subproblem in each iteration}
    \begin{align}
    \underset{\pb}{\mathrm{minimize}} & \sum\nolimits_{i=1}^{\Ns}\hat{\Jb}(i,i)p_i-2\sum\nolimits_{i=1}^{\Ns}\Re\big\{\hat{\Kb}(i,i) \sqrt{p_i}\big\}  \\
    \mathrm{subject~to} & \sum\nolimits_{i=1}^{\Ns}p_i\leq \Pt,
\end{align}
\end{subequations}
with $\hat{\Jb}  = \Vb^\H\Hb^\H\big( \Gb \hat{\Sb}\Gb+ \diag( \hat{\Sb})\Gb(\Ib-\Gb) \big)\Hb \Vb$ and $ \hat{\Kb}  = \diag(\hat{\pb})\Vb^\H\Hb^\H\Gb \hat{\Cb}_{\eb}^{-1}\Gb \Hb \Vb \diag(\hat{\pb})$.
Setting the derivative of the Lagrangian of \eqref{pb:simplified subproblem in each iteration} to zero, we obtain the solution as
\begin{equation}\label{eq:closed-form q}
    p_i=\sqrt{\frac{\hat{\Kb}(i,i)}{\hat{\Jb}(i,i)+\mu}}, \quad \forall i \, ,
\end{equation}
where $\mu$ is the Lagrange multiplier and can be computed via a bisection search over $(0,\mu_{\rm ub}]$ to satisfy $\sum_{i=1}^{\Ns}p_i= \Pt$. Here, $\mu_{\rm up}= \frac{1}{\sqrt{\Pt}} \big\|\diag(\hat{\Kb})\big\|_{\rm F}$ is obtained by setting $\hat{\Jb}=\mathbf{0}$. 

The proposed digital precoding design is summarized in Algorithm~\ref{alg:Type-II DBF Design}. We note that the conventional WF solution, denoted by $\Fb_{\rm WF}$, is obtained by assuming $\Cb_{\etab}=\mathbf{0}$ as in full-resolution systems. Hence, $\Fb_{\rm WF}$ is sub-optimal for \eqref{pb:simplified subproblem in each iteration} due to the presence of the QD, as will be shown in Section~\ref{sec_simulation}.

\begin{algorithm}[t]
\footnotesize
\caption{MM-based digital precoding design}\label{alg:Type-II DBF Design}
\LinesNumbered 
\KwOut{$\Fb$}
Initialize $\pb, \epsilon $.\\
\Repeat{ $\left|R(\pb)-R(\hat{\pb}) \right|\leq \epsilon  $}{
$\hat{\pb} \leftarrow \pb$.\\
Update $\pb$ as in \eqref{eq:closed-form q}.\\
 }
Obtain $\Fb$ as in \eqref{eq:SCA precoder}.
\end{algorithm}

\subsection{Hybrid Precoding Design}
Define $f(\Frf,\Fbb)= \tr(\hat{\Lb}\Frf\Fbb\Fbb^\H \Frf^\H)-2\Re\big\{\tr( \hat{\Db} \Frf\Fbb  )\big\}$, with $\hat{\Db} = \hat{\Xb}^\H\hat{\Cb}_{\eb}^{-1}\Gb\Hb$ and $\hat{\Lb} = \Hb^\H \big( \diag(\hat{\Sb})\Gb(\Ib-\Gb)+\Gb\hat{\Sb}\Gb \big)\Hb$.
Using Lemma~\ref{thrm:log-det MM surrogate function}, hybrid precoders can be designed in each iteration as
\begin{subequations}\label{pb:hybrid precoding problem}
    \begin{align}
        \underset{\Fb,\Frf,\Fbb}{\mathrm{minimize}}\ & f(\Frf,\Fbb) \\
        \mathrm{subject~to}\ & \text{ \eqref{eq: coupling constraints} and \eqref{eq: analog components}}, \nonumber \\
        & \|\Fb\|_{\rm F}^2 \leq \Pt,
    \end{align}
\end{subequations}
which is challenging due to \eqref{eq: coupling constraints} and \eqref{eq: analog components}. Observing that $f(\Frf,\Fbb)$ is convex with respect to $\Frf$ (resp. $\Fbb$) when $\Fbb$ (resp. $\Frf$) is fixed, we compute the analog and digital precoders via alternating optimization, as explained next.

\textbf{Update $\Frf$:} For a given $\Fbb$, $\Frf$ can be designed via $\mathrm{minimize}_{\Frf \in \Fcl} \; f(\Frf,\Fbb) $, with
$\Fcl=\Fcl_{\rm full}$ for FC-HBF and $\Fcl=\Fcl_{\rm sub}$ for PC-HBF. To efficiently solve this problem, we propose to use the projected gradient descent (PGD) method. Specifically, let $\nabla_{\Frf} f =2 (\hat{\Lb} \Frf \Fbb -\hat{\Db}^\H) \Fbb^\H$ denote the gradient of $f(\Frf,\Fbb)$ with respect to $\Frf$.
Let ${\rm Proj}_{\Fcl}(\Frf)$ represent the operation that projects $\Frf$ onto $\Fcl$. Define $\Wb= \blkdiag(\mathbf{1}_M,\ldots,\mathbf{1}_M)$, where $\mathbf{1}_M \in \Rs^M$ is an all-one vector. The projectors for FC-HBF and PC-HBF are given by ${\rm Proj}_{\Fcl_{\rm full}}(\Frf) \!=\!e^{j\angle \Frf}$ and ${\rm Proj}_{\Fcl_{\rm sub}}(\Frf)\! =\!\Wb \! \odot e^{j\angle \Frf}$,
respectively, where $\angle \Frf$ returns the angles of each entry of $\Frf$ and $\odot$ denotes the Hadamard product. The update rules for the PGD method can be expressed as
\begin{equation}\label{eq:PGD update rules}
    \Frf\leftarrow {\rm Proj}_{\Fcl}\left(\tilde{\Fb}_{\rm RF}-\beta  \frac{ \nabla_{\Frf}f }{\| \nabla_{\Frf} f\|_{\rm F}}\bigg|_{\Frf=\tilde{\Fb}_{\rm RF}} \right),
\end{equation}
where $\beta$ denotes the step size and $\tilde{\Fb}_{\rm RF}$ is the iterate of $\Frf$. The detailed steps for obtaining $\Frf$ are similar to those in \cite[Alg.~1]{ma2024hybrid} and are thus omitted.

\textbf{Update $\Fbb$:} For a given $\Frf$,  we can derive a closed-form solution for $\Fbb$ as 
\begin{equation}\label{eq:Fbb closed-form}
    \Fbb=( \Frf^\H\hat{\Lb}\Frf + \lambda \Frf^\H \Frf)^{-1}\Frf^\H\hat{\Db}^\H, 
\end{equation}
where $\lambda$ is the Lagrange multiplier associated with the power constraint and can be computed via a bisection search over $(0, \lambda_{\rm ub}]$, with $\lambda_{\rm ub} = \frac{1}{\sqrt{\Pt}} \| \left(\Frf^\H\Frf \right)^{-\frac{1}{2}} \Frf^\H\hat{\Db}^\H\|_{\rm F}$ obtained from the complementary slackness condition $\lambda\left(\tr\left(\Frf^\H\Frf\Fbb \Fbb^\H \right)-\Pt \right)=0$. 
 
The proposed hybrid precoding design is summarized in Algorithm~\ref{alg:AltMin for FC-HBF}. The precoder $\Fb$ is initialized with the conventional WF precoder $\Fb_{\rm WF}$. Let $\tilde{\Vb}$ contain the $\Nrf$ right singular vectors of $\Hb$ associated with the $\Nrf$ largest singular values. We initialize $\Frf=\tilde{\Vb}$ (resp. $\Frf=\Wb \odot \tilde{\Vb}$) for FC-HBF (resp. PC-HBF), which yields $\Fbb=\Frf^{\dagger} \Fb_{\rm WF} $. Algorithms~\ref{alg:Type-II DBF Design} and~\ref{alg:AltMin for FC-HBF} are guaranteed to converge according to the MM theory.

\begin{algorithm}[t]
\footnotesize
\caption{MM-based hybrid precoding design}\label{alg:AltMin for FC-HBF}
\LinesNumbered 
\KwOut{$\Frf,\Fbb$}
Initialize $\Frf\in\Fcl,\Fb$, $\Fbb=\Frf^{\dagger}\Fb,\epsilon $.\\
\Repeat{ $\left|R(\Fb)-R(\hat{\Fb}) \right|\leq \epsilon  $}{
$\hat{\Fb}\leftarrow \Fb $ .\\
\Repeat{$f(\Frf,\Fbb)$ stops decreasing}
{
Update $\Frf$ via the PGD method.\\
Update $\Fbb$ as in \eqref{eq:Fbb closed-form}.
}
$\Fb=\Frf \Fbb$.
 }
\end{algorithm}

\subsection{Complexity Analysis}\label{eq:complexity analysis}
Let $I_1$ be the number of iterations for Algorithm~\ref{alg:Type-II DBF Design}. With $\Ns \ll \min(\Nr,\Nt)$, the complexity of Algorithm~\ref{alg:Type-II DBF Design} can be expressed as $I_1\Ocl\left( 2\Nr^2\Nt +2\Nr^3+ 4\Nr^2\Ns+ 6\Nr\Nt\Ns\right)$ floating-point operations (FLOPs) due mainly to matrix multiplications and inverses. For Algorithm~\ref{alg:AltMin for FC-HBF}, step~5 has a complexity of $ \Ocl(I_{\rm pgd}\Nt^2\Nrf)$~FLOPs due to the matrix multiplication, where $I_{\rm pgd}$ denotes the number of PGD iterations, whereas computing $\Fbb$ in step~6 requires $\Ocl(\Nt^2\Nrf)$ FLOPs. Furthermore, $\hat{\Xb}$, $\hat{\Cb}_{\eb}$, $\hat{\Lb}$, and $\hat{\Db}$ are computed in the outer loop, requiring $\Ocl\left( 2\Nr^2\Nt +2\Nr^3+ 6\Nr^2\Nrf+ 5\Nr\Nt\Nrf\right)$~FLOPs.~Therefore, the overall complexity of Algorithm~\ref{alg:AltMin for FC-HBF} can be~expressed as~$I_{\rm out}\Nrf \Ocl\left(6\Nr^2+ 5\Nr\Nt\!\!+\!I_{\rm in}I_{\rm pgd}\Nt^2 \right)+I_{\rm out}\Ocl\left( 2\Nr^2\Nt +2\Nr^3\right)$~FLOPs, where $I_{\rm in}$ and $I_{\rm out}$ denote, respectively, the number of iterations for the inner and outer loops of Algorithm~\ref{alg:AltMin for FC-HBF}. The complexity of the proposed algorithms is summarized and compared with that in \cite{mo2017hybrid,yu2016alternating,gao2016energy,ma2024joint} in Table~\ref{tb:Complexity of proposed algorithms}. We empirically observe from simulations that setting $I_{\rm in}=I_{\rm pgd}=1$ has negligible effects on the convergence and performance of Algorithm~\ref{alg:AltMin for FC-HBF}. Furthermore, the proposed algorithms have similar
convergence speed, with a few hundreds of iterations required for convergence.

\medskip

\section{Numerical Results}\label{sec_simulation}
In the simulations, we set $\Nt=\Nr=64$ and $\Ns=\Nrf=8$, and consider $1~\ghz$ bandwidth. The Saleh-Valenzuela channel model is employed with channel parameters configured as in \cite{yu2016alternating}. The signal-to-noise ratio (SNR) is defined as $\SNR = \frac{\Pt}{\sigma_{\rm n}^2}$. Unless otherwise stated, we assume $\SNR=20~\dB$ and adopt identical $b$-bit ADCs for each RF chain. Other parameters are detailed in the caption of each figure.

For comparison, we consider the following baselines:
\begin{itemize}
    \item ``UqOpt'': Using $\Fb_{\rm WF}$ for maximizing the SE with full-resolution ADCs.
    \item ``DBF: WF'': Using $\Fb_{\rm WF}$ with low-resolution ADCs.
    \item ``DBF: JPC'': Joint design of digital precoder and combiner for low-resolution ADCs \cite{ma2024joint}.
    \item ``FC-HBF: SVD-based'': SVD-based FC hybrid precoding with low-resolution ADCs \cite{mo2017hybrid}.
    \item ``FC-HBF: AO'': FC hybrid precoding design for full-resolution systems \cite{yu2016alternating}.
    \item ``PC-HBF: SVD-based'': Modified SVD-based design \cite{mo2017hybrid} with projection of the analog precoder onto $\Fcl_{\rm sub}$.
    \item ``PC-HBF: SIC'': PC hybrid precoding design for full-resolution systems \cite{gao2016energy}.
\end{itemize}
The numerical results for all the precoding schemes are obtained by averaging over $10^3$ independent channel realizations and are based on the simulated $\Cb_{\etab}$ for a more practical performance characterization (rather than on the diagonal approximation of $\Cb_{\etab}$ used in \eqref{eq:CoV eff approx}). The simulated $\Cb_{\etab}$ is obtained by the scaling law of optimal quantization \cite[Prop.~1]{ma2024joint} and with sample average over $10^5$~realizations.

  \begin{figure*}[t]
\small
\vspace{-0.3cm}
    \centering
    \hspace{-2mm}
    \subfigure[SE versus SNR ($b=3$).]
    {\label{fig:SE_BitSNR0Nt16Nr36Ns4_apx}\includegraphics[width=0.32\textwidth]{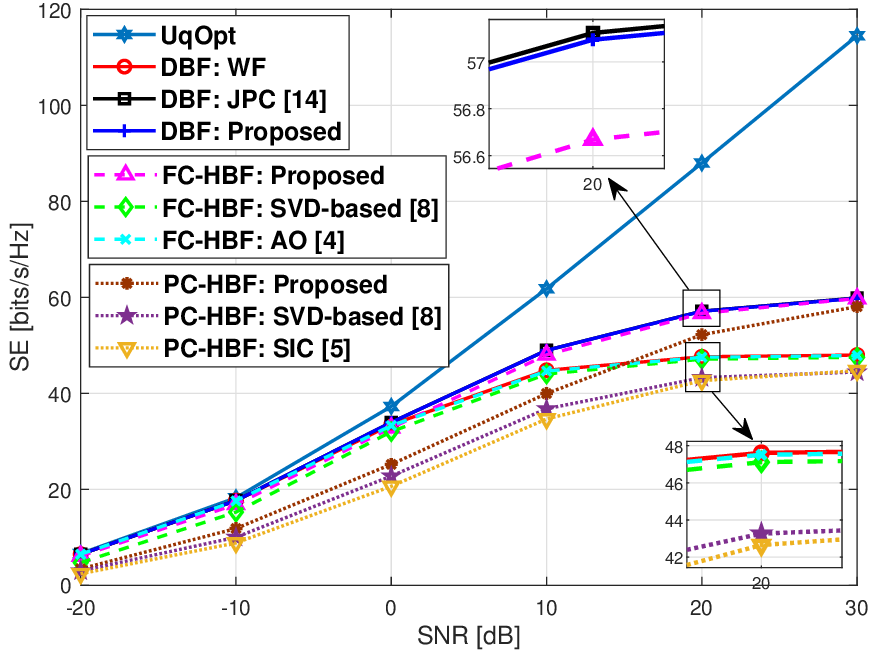}}
    \subfigure[SE versus ADC resolution.]
    {\label{fig:SE_BitSNR30Nt16Nr36Ns4_apx} \includegraphics[width=0.32\textwidth]{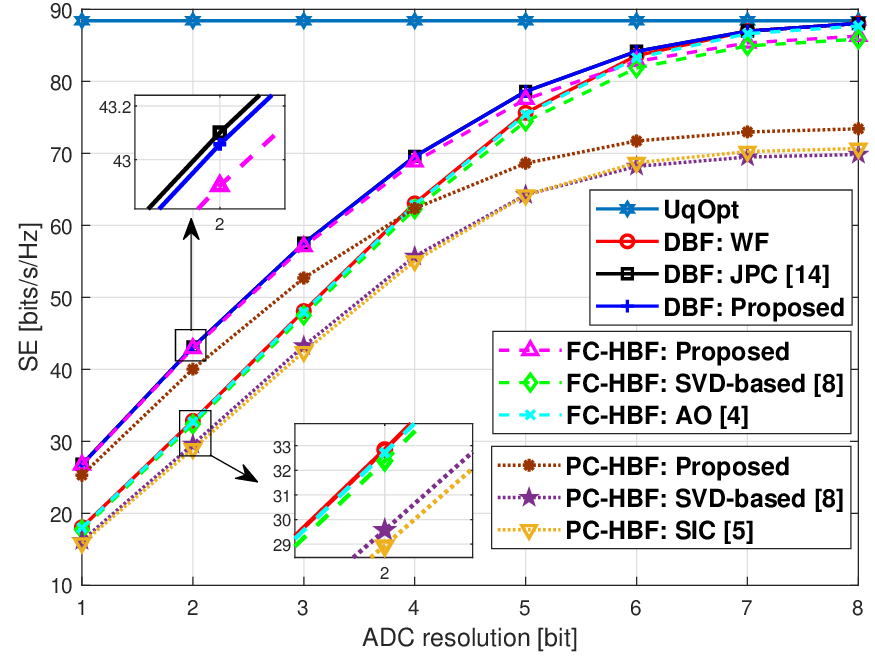}}
        \subfigure[SE versus $\xi$ ($b=3$).]
    {\label{fig:SE_CSI} \includegraphics[width=0.32\textwidth]{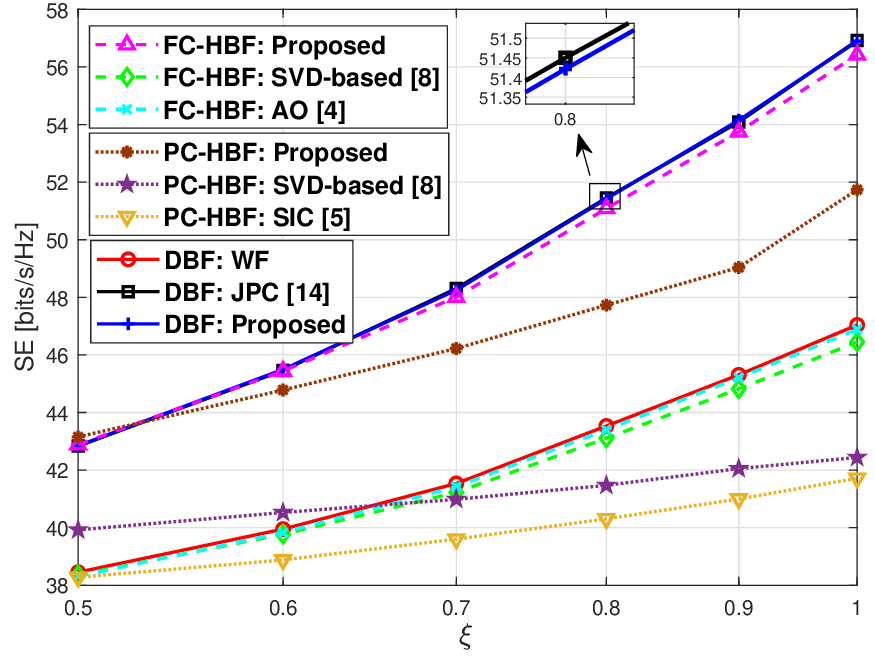}}
    \vspace{-2mm}
    \caption{SE performance. The ``DBF: Proposed'' corresponds to Algorithm~\ref{alg:Type-II DBF Design}, while  the ``FC-HBF: Proposed'' and ``PC-HBF: Proposed'' represents Algorithm~\ref{alg:AltMin for FC-HBF} with projectors ${\rm Proj}_{\Fcl_{\rm full}}(\Frf)$ and ${\rm Proj}_{\Fcl_{\rm sub}}(\Frf)$ in the PGD, respectively.}
    \label{fig:SE performance} 
    \vspace{-4mm}
\end{figure*}

\begin{table}[t]
\footnotesize
    \caption{Computational complexity of the considered algorithms. Here, $I_{\rm n}$ and $I_{\rm c}$ denote the number of iterations of the designs in \cite{ma2024joint} and \cite{mo2017hybrid}, respectively, while $I_{\rm o}$ and $I_{\rm i}$ represent the number of iterations of the outer and inner loop for the method in \cite{yu2016alternating}.} \label{tb:Complexity of proposed algorithms}
\centering
\renewcommand{\arraystretch}{1.2}

\begin{threeparttable}
    \begin{tabular}{|c|c|} 
    \hline 
      \textbf{Algorithm} & \textbf{Computational complexity}\\
          \hline
          \hline
     Algorithm~\ref{alg:Type-II DBF Design} & $I_1\Ocl\left( 2\Nr^2\Nt +2\Nr^3+ 4\Nr^2\Ns+ 6\Nr\Nt\Ns\right)$\\
       \hline 
             DBF \cite{ma2024joint} &  $I_{\rm n}\Ocl(3\Nt^3+3\Nt^3+8\Nt^2\Nr+8\Nt\Nr^2)$  \\
     \hline 
    \hline  
    Algorithm~\ref{alg:AltMin for FC-HBF} & \makecell[c]{$I_{\rm out}\Nrf \Ocl\left(6\Nr^2+ 5\Nr\Nt+I_{\rm in}I_{\rm pgd}\Nt^2 \right) $\\$+I_{\rm out}\Ocl\left( 2\Nr^2\Nt +2\Nr^3\right)$}  \\
    \hline 

    FC-HBF \cite{mo2017hybrid}  &  $\Ocl(\Nt\Nr\Nrf+ I_{\rm c}\Nt\Nrf^2+ 2I_{\rm c}\Nrf^3)$  \\
        \hline 
    FC-HBF \cite{yu2016alternating}  & $I_{\rm o}\Ocl\left( 2\Nt\Nrf^2 + I_{\rm i}\Nt^2\Nrf \right)$   \\
    \hline 
     PC-HBF \cite{gao2016energy}  &  $\Ocl\left( \Nt^2\Nrf + 2\Nr\Nt \Nrf \right)$\\
    \hline
    \end{tabular}
    \end{threeparttable}
\end{table}

Figs.~\ref{fig:SE_BitSNR0Nt16Nr36Ns4_apx} and~\ref{fig:SE_BitSNR30Nt16Nr36Ns4_apx} plot the SE versus the SNR and ADC resolution, respectively. From Fig.~\ref{fig:SE_BitSNR0Nt16Nr36Ns4_apx}, we observe that the proposed digital precoding design achieves an SE comparable to that of ``DBF: JPC''. Furthermore, the proposed FC-HBF and PC-HBF schemes outperform their benchmark counterparts, especially at high SNR. For example, at $\SNR=20$~dB, the proposed FC-HBF and PC-HBF approaches achieve $19\%$ and $21\%$ improvements in SE over ``FC-HBF: AO'' and ``PC-HBF: SVD-based'', respectively. Additionally, both the proposed FC and PC hybrid precoding designs outperform the ``DBF: WF'' in low-resolution systems at high SNR. In particular, the proposed FC hybrid precoding design attains an SE comparable to that of the proposed digital one. Fig.~\ref{fig:SE_BitSNR30Nt16Nr36Ns4_apx} shows that the proposed DBF, FC-HBF, and PC-HBF schemes achieve significantly higher SE compared with their benchmarks, especially for low ADC resolutions. Notably, PC-HBF achieves an SE comparable to those of DBF and FC-HBF as the ADC resolution decreases. 

Fig.~\ref{fig:SE_CSI} shows the impact of imperfect CSI on the proposed designs with $b=3$. The estimated channel matrix $\hat{\Hb}$ is modeled as $\hat{\Hb}=\xi \Hb + \sqrt{1-\xi^2}\Eb$ \cite{jacobsson2019linear}
where $\Hb$ denotes the true channel, $\xi \in [0,1]$ controls the channel estimation accuracy, and $\Eb$ represents the estimation error with entries drawn from the distribution $\mathcal{CN}(0,1)$. It is observed from Fig.~\ref{fig:SE_CSI} that the imperfect CSI affects all the proposed and baseline schemes similarly, resulting in comparable levels of SE degradation. Nonetheless, with higher CSI accuracy, the proposed schemes demonstrate greater SE gains over the baselines, validating their efficiency.

\begin{figure}[t]
\small
    \centering
    \hspace{-2mm}
    \subfigure[EE performance.]
    {\label{fig:EE cmp}\includegraphics[width=0.24\textwidth]{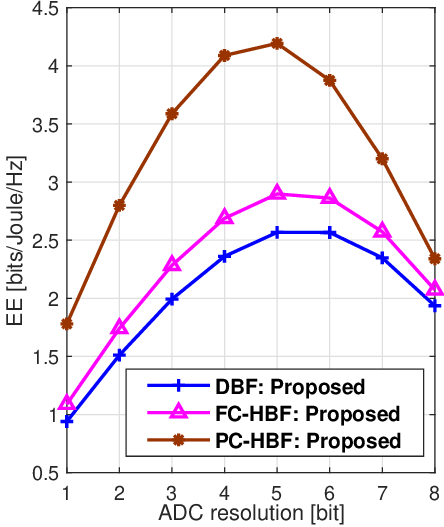}}
     \hspace{-2mm} 
    \subfigure[Execution time comparison.]
    {\label{fig:TC_Bit} \includegraphics[width=0.24\textwidth]
    {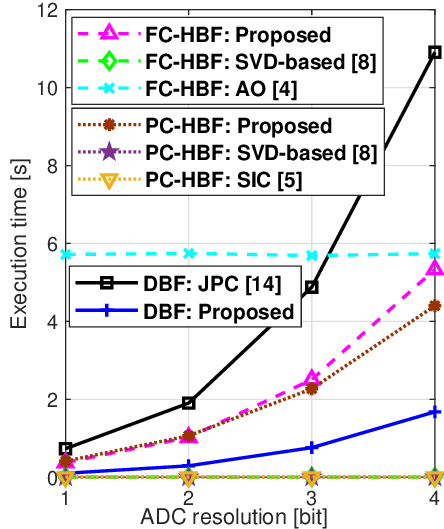}}
    \vspace{-2mm}
    \caption{EE and execution time versus ADC resolution.}
    \label{fig:EE and TC} 
    \vspace{-1mm}
\end{figure}

Fig.~\ref{fig:EE cmp} plots a comparison of the EE, given by $\frac{R}{P_{\rm T}+P_{\rm R}}$, where $P_{\rm T}, P_{\rm R}$ respectively represent the power consumption of the Tx and Rx \cite{gao2016energy,nguyen2019unequally,mo2017hybrid}. Specifically, we assume $P_{\rm R}= \Nr\left( P_{\rm LNA} + P_{\rm RF}+ 2P_{\rm ADC} \right)$, where $P_{\rm LNA}$, $P_{\rm ADC}$, and $P_{\rm RF}$ denote the power consumption of the low-noise amplifier, the ADC, and the remaining components of the RF chain, respectively. Furthermore, $P_{\rm T} \in \{P_{\text{ DBF}}, P_{\text{FC-HBF}}, P_{\text{PC-HBF}}\}$ where $P_{\text{ DBF}}$, $P_{\text{FC-HBF}}$, and $P_{\text{PC-HBF}}$ denote the Tx power consumption of the DBF, FC-HBF, and PC-HBF architectures, respectively, determined similarly to the definitions in \cite{gao2016energy,nguyen2019unequally,mo2017hybrid}. The ADC power consumption is typically modeled as $P_{\rm ADC}=\kappa \fs 2^b$, where $\kappa$ and $\fs$ respectively denote the figure of merit and the sampling frequency (ideally equal to the system bandwidth) \cite{murmann2015race}. In the following simulations, we set $P_{\rm RF}=43~\mW$, $P_{\rm LNA}=25~\mW$, and $\kappa=494~{\rm fJ/step/Hz}$ \cite{ma2024hybrid}. It is seen from Fig.~\ref{fig:EE cmp} that the PC-HBF architecture achieves over $1.5\times$ higher EE than DBF and FC-HBF with few-bit ADCs. Additionally, using a moderate number of ADC bits per RF chain significantly improves the EE across all three beamforming architectures.

Fig.~\ref{fig:TC_Bit} depicts the average execution time required for the algorithms for ADC resolutions up to $4$ bits with Intel(R) Xeon(R) Gold 6226 CPUs. We observe that the proposed designs are more time-efficient than ``DBF: JPC'' \cite{ma2024joint} and ``FC-HBF: AO'' \cite{yu2016alternating} while requiring more execution time than the ``FC-HBF: SVD-based'' \cite{mo2017hybrid} and ``PC-HBF: SIC''  \cite{gao2016energy}. The results align with the differences in complexity of the algorithms, as shown in Table~\ref{tb:Complexity of proposed algorithms}.

\section{Conclusions}
We investigated precoding designs for DBF, FC-HBF, and PC-HBF architectures in massive MIMO systems with low-resolution ADCs to maximize the system's SE. To solve the resulting challenging nonconvex problem, we first derived a tight lower bound for the SE and then iteratively optimized 
the precoder under the MM framework. Based on this method, we proposed efficient precoding algorithms for DBF, FC-HBF, and PC-HBF architectures. Numerical results demonstrated that the proposed methods significantly outperform the compared benchmarks in low-resolution systems, especially with 1-bit ADCs. Furthermore, the results indicated that the PC hybrid precoding design can achieve SE comparable to those of the digital and FC hybrid precoding designs in 1-bit systems, highlighting the potential of PC-HBF architectures. For future work, we will explore digital and hybrid precoding for multiuser massive MIMO with low-resolution ADCs.


\bibliographystyle{IEEEtran}
\bibliography{conf_short,jour_short,refs-my}

\end{document}

%% file: preamble.tex
\usepackage{diagbox} 
\usepackage{graphicx}
\usepackage{epstopdf}
\usepackage{psfrag}
\usepackage{subfigure}
\usepackage{url}
\usepackage{stfloats}
\usepackage{amsfonts,amssymb,amsmath,bm,paralist,theorem,cite,ifthen,color,nccmath}
\usepackage[labelsep=period,font=small]{caption}
\usepackage{calc}
\usepackage{enumerate}
\usepackage{multirow}
\usepackage{makecell}
\usepackage[ruled]{algorithm2e}
\usepackage{setspace}
\usepackage{array}
\usepackage{float}
\usepackage{soul}
\usepackage[flushleft]{threeparttable}

\usepackage[colorlinks,
            linkcolor=blue,
            anchorcolor=red,
            citecolor=red]{hyperref}
\usepackage[]{mdframed}

\SetKwInOut{Input}{Input}
\SetKwInOut{Output}{Output}

\newcommand{\T}{{\scriptscriptstyle\mathsf{T}}}
\renewcommand{\H}{{\scriptscriptstyle\mathsf{H}}}

\newtheorem{lemma}{Lemma}

\graphicspath{{Figures/}}

\newcounter{algoline}

\AtBeginEnvironment{algorithm}{\setcounter{algoline}{0}}

\newcommand\Ccl{\ensuremath{\mathcal{C}}}

\newcommand\Ncl{\ensuremath{\mathcal{N}}}
\newcommand\Ocl{\ensuremath{\mathcal{O}}}

\newcommand\Tcl{\ensuremath{\mathcal{T}}}
\newcommand\Fcl{\ensuremath{\mathcal{F}}}

\newcommand\Cs{\ensuremath{{\mathbb{C}}}}
\newcommand\Es{\ensuremath{{\mathbb{E}}}}
\newcommand\Rs{\ensuremath{{\mathbb{R}}}}

\newcommand\Ab{\ensuremath{ \mathbf{A} }}

\newcommand\Cb{\ensuremath{ \mathbf{C} }}
\newcommand\Db{\ensuremath{ \mathbf{D} }}
\newcommand\Eb{\ensuremath{ \mathbf{E} }}
\newcommand\Fb{\ensuremath{ \mathbf{F} }}
\newcommand\Gb{\ensuremath{ \mathbf{G} }}
\newcommand\Hb{\ensuremath{ \mathbf{H} }}
\newcommand\Ib{\ensuremath{ \mathbf{I} }}

\newcommand\Vb{\ensuremath{ \mathbf{V} }}
\newcommand\Wb{\ensuremath{ \mathbf{W} }}
\newcommand\Jb{\ensuremath{ \mathbf{J} }}
\newcommand\Kb{\ensuremath{ \mathbf{K} }}
\newcommand\Lb{\ensuremath{ \mathbf{L} }}

\newcommand\Sb{\ensuremath{ \mathbf{S} }}
\newcommand\Xb{\ensuremath{ \mathbf{X} }}

\newcommand\ab{\ensuremath{ \mathbf{a} }}

\newcommand\eb{\ensuremath{ \mathbf{e} }}

\newcommand\fb{\ensuremath{ \mathbf{f} }}

\newcommand\nb{\ensuremath{ \mathbf{n} }}

\newcommand\pb{\ensuremath{ \mathbf{p} }}

\newcommand\ssb{\ensuremath{ \mathbf{s} }}

\newcommand\yb{\ensuremath{ \mathbf{y} }}
\newcommand\zb{\ensuremath{ \mathbf{z} }}

\newcommand\Gammab{\ensuremath{{\bm \Gamma}}}

\newcommand\etab{\ensuremath{{\bm \eta}}}

\newcommand\diag{\ensuremath{{\rm diag}}}
\newcommand\tr{\ensuremath{{\rm Tr}}}

\newcommand{\SNR}{\textrm{SNR}}
\newcommand\ghz{\textrm{GHz}}

\newcommand\dB{\textrm{dB}}
\newcommand\mW{\textrm{mW}}

\newcommand\Frf{\ensuremath{ \mathbf{F}_{\rm RF} }}

\newcommand\Fbb{\ensuremath{ \mathbf{F}_{\rm BB} }}

\newcommand\Nt{\ensuremath{ N_{\rm t} }}
\newcommand\Nr{\ensuremath{ N_{\rm r} }}

\newcommand\Ns{\ensuremath{ N_{\rm s} }}
\newcommand\Nrf{\ensuremath{ N_{\rm RF} }}

\newcommand\fs{\ensuremath{ f_{\rm s} }}

\newcommand\Pt{\ensuremath{ P_{\rm t} }}

\newcommand\sigman{\ensuremath{ \sigma_{\rm n} }}

\newcommand\blkdiag{\ensuremath{{\rm blkdiag}}}

\usepackage{relsize}

 \usepackage{afterpage}

\usepackage{etoolbox}
\usepackage{titlesec}
\titlespacing{\section}{-0.64 cm}{2pt}{2pt}
\titlespacing{\subsection}{0 cm}{2pt}{2pt}
\usepackage[normalem]{ulem}